\documentclass[twocolumn]{aastex63}
\newcommand{\alopeke}{`Alopeke}
\newcommand{\PS}{Pan-STARRS1}

\newcommand{\mardust}{A_r}  % Galactic extinction in the r-band

\newcommand{\maidust}{A_i}  % Galactic extinction in the i-band

\newcommand{\refstara}{Star-1}
\newcommand{\refstarb}{Star-2}

\newcommand{\frbname}{FRB\,20180916B}
\newcommand{\frb}{FRB\,20180916B}
\newcommand{\frbeventa}{FRB\,20201023}  % Our event's name
\newcommand{\frbeventb}{FRB\,20220908}  % Our event's name

\newcommand{\period}{16.3} % in days (approx)

\newcommand{\stararmag}{15.7481\pm0.0030}  % Updated from Panstarrs DR2
\newcommand{\staraimag}{15.1387\pm0.0069} % Updated from Panstarrs DR2
\newcommand{\starbrmag}{17.1016\pm0.0225}  % Updated from Panstarrs DR2
\newcommand{\starbimag}{16.6247\pm0.0121} % Updated from Panstarrs DR2

\newcommand{\chimearrivala}{UTC 2020-10-23 07:48:30.778}  % From Tendulkar. Need to calibrate to a well defined relative frame (e.g. barycenter?) 
\newcommand{\opticalarrivalCHIMEa}{UTC 2020-10-23T07:48:21.695}
\newcommand{\opticalarrivalAlopekea}{UTC 2020-10-23T07:48:21.709}
\newcommand{\chimearrivalb}{UTC 2022-09-08T10:53:26.889}  % From Tendulkar. Need to calibrate to a well defined relative frame (e.g. barycenter?) 
\newcommand{\opticalarrivalCHIMEb}{UTC 2022-09-08T10:53:17.817}
\newcommand{\opticalarrivalAlopekeb}{UTC 2022-09-08T10:53:17.831}
\newcommand{\mjdeventa}{59145.32525072763}
\newcommand{\mjdeventb}{59830.45367810875}

\newcommand{\aata}{$160 \pm 0.07$}  % in ms ; from S. Howell email Sep 15, 2020, 6:19 PM 
\newcommand{\dtshort}{10.4}
\newcommand{\dtres}{10.419} % on-sky exposure time, ms
\newcommand{\dttot}{11.595} % total time between frames, ms

\newcommand{\mcunit}{{\rm e^{-}~exposure^{-1}}}
\newcommand{\cunit}{$\mcunit$}
\newcommand{\cstara}{\ensuremath{C_{\rm Star-1}^{\rm Tot}}}
\newcommand{\cstarb}{\ensuremath{C_{\rm Star-2}^{\rm Tot}}}
\newcommand{\cbkg}{\ensuremath{C_{\rm bkg}}}
\newcommand{\ctfrb}{\ensuremath{C_{\rm FRB}^{\rm Tot}}} % Counts at FRB location outside the event
\newcommand{\mcofrb}{C_{\rm FRB}^{\rm other}} % Counts at FRB location  outside the event
\newcommand{\cofrb}{$\mcofrb$}
\newcommand{\mcfrb}{C_{\rm FRB}^{\rm FRB}} %  Counts during FRB event
\newcommand{\cfrb}{$\mcfrb$}
\newcommand{\mmxfrb}{{\rm max}(\ctfrb)}
\newcommand{\mxfrb}{$\mmxfrb$}

\newcommand{\mmufrb}{\mu_{\rm FRB}} % Mean counts for the FRB
\newcommand{\mufrb}{$\mmufrb$}
\newcommand{\mumufrb}{\mu_{\rm FRB}^{99.9}} % Upper limit

\newcounter{daggerfootnote}

\usepackage{amsmath}
\usepackage{color}
\usepackage{textcomp}

\begin{document}

\title{Limits on Optical Counterparts to the Repeating \frbname\ from High-speed Imaging with Gemini-N/\alopeke}

\correspondingauthor{Charles D. Kilpatrick; Nicolas Tejos}
\email{ckilpatrick@northwestern.edu; nicolas.tejos@pucv.cl}

\newcommand{\PUCV}{\affiliation{
Instituto de F\'{i}sica, Pontificia Universidad Cat\'{o}lica de Valpara\'{i}so, Casilla 4059, Valpara\'{i}so, Chile}}

\newcommand{\NU}{\affiliation{Center for Interdisciplinary Exploration and Research in Astrophysics (CIERA) and Department of Physics and Astronomy, Northwestern University, Evanston, IL 60208, USA}}

\newcommand{\UCSC}{\affiliation{Department of Astronomy and Astrophysics, University of California, Santa Cruz, CA 95064, USA}}

\newcommand{\NMSU}{\affiliation{Astronomy Department, Box 30001, Department 4500, New Mexico State University, Las Cruces, NM 88003-0001}}

\newcommand{\APO}{\affiliation{Apache Point Observatory and New Mexico State University, P.O. Box 59, Sunspot, NM, 88349-0059, USA}}

\newcommand{\UCB}{\affiliation{Astronomy Department and Theoretical Astrophysics Center, University of California, Berkeley, Berkeley, CA 94720, USA}}

\newcommand{\UI}{\affiliation{Centre for Astrophysics and Cosmology, Science Institute, University of Iceland, Dunhagi 5, 107 Reykjav\'{i}k, Iceland}}

\newcommand{\CGWC}{\affiliation{Center for Gravitational Waves and Cosmology, West Virginia University, Chestnut Ridge Research Building, Morgantown, WV 26505, USA}}

\newcommand{\WVU}{\affiliation{Department of Physics and Astronomy, West Virginia University, P.O. Box 6315, Morgantown, WV 26506, USA}}

\newcommand{\dunlap}{\affiliation{Dunlap Institute for Astronomy \& Astrophysics, University of Toronto, 50 St. George Street, Toronto, ON M5S 3H4, Canada}}

\newcommand{\toronto}{\affiliation{David A. Dunlap Department of Astronomy \& Astrophysics, University of Toronto, 50 St. George Street, Toronto, ON M5S 3H4, Canada}}

\newcommand{\ucscscience}{\affiliation{Division of Physical and Biological Sciences, University of California Santa Cruz, 1156 High Street, Santa Cruz, CA 95064, USA}}

\newcommand{\ames}{\affiliation{NASA Ames Research Center, Moffett Field, CA 94035, USA}}

\newcommand{\gemini}{\affiliation{Gemini Observatory/ NSF's NOIRLab, 670 A'ohoku Place, Hilo, HI, 96720, USA}}

\newcommand{\tifr}{\affiliation{Department of Astronomy and Astrophysics, Tata Institute of Fundamental Research, Mumbai, 400005, India}}

\newcommand{\ncra}{\affiliation{National Centre for Radio Astrophysics, Post Bag 3, Ganeshkhind, Pune, 411007, India}}

\newcommand{\cifar}{\affiliation{CIFAR Azrieli Global Scholars Program, MaRS Centre, West Tower, 661 University Ave, Suite 505, Toronto, ON, M5G 1M1 Canada}}

\author[0000-0002-5740-7747]{Charles D. Kilpatrick}
\NU

\author[0000-0002-1883-4252]{Nicolas Tejos}
\PUCV

\author[0000-0002-7738-6875]{J. Xavier Prochaska}
\UCSC

\author{Consuelo N\'u\~nez}
\PUCV

\author[0000-0001-8384-5049]{Emmanuel Fonseca}
\CGWC\WVU

\author[0000-0003-4236-6927]{Zachary Hartman} % Alopeke
\gemini

\author[0000-0002-2532-2853]{Steve B. Howell} % Alopeke
\ames

\author[0000-0001-5605-1702]{Tom Seccull} % Alopeke
\gemini

\author[0000-0003-2548-2926]{Shriharsh P. Tendulkar} % proposal 2020B
\tifr\ncra\cifar

\shorttitle{\frbname\ High-speed optical limits}
\shortauthors{Kilpatrick et al.}

\received{\today}
\revised{\today}
\accepted{\today}

%% Command to document which AAS Journal the manuscript was submitted to.
%% Adds "Submitted to " the argument.

\submitjournal{ApJ}

\begin{abstract}
We report on contemporaneous optical observations at $\approx 10$\,ms timescales from the fast radio burst (FRB) 20180916B of two repeat bursts (\frbeventa, \frbeventb) taken with the \alopeke\ camera on the Gemini North telescope.  These repeats have radio fluences of 2.8 and 3.5~Jy~ms, respectively, approximately in the lower 50th percentile for fluence from this repeating burst. The \alopeke\ data reveal no significant optical detections at the FRB position and we place $3\sigma$ upper limits to the optical fluences of $<8.3\times10^{-3}$ and $<7.7\times10^{-3}$~Jy~ms after correcting for line-of-sight extinction. Together, these yield the most sensitive limits to the optical-to-radio fluence ratio of an FRB on these timescales with $\eta_{\nu} < 3\times10^{-3}$ by roughly an order of magnitude. These measurements rule out progenitor models where \frbname\ has a similar fluence ratio to optical pulsars similar to the Crab pulsar or optical emission is produced as inverse Compton radiation in a pulsar magnetosphere or young supernova remnant. Our ongoing program with \alopeke\ on Gemini-N will continue to monitor repeating FRBs, including \frbname, to search for optical counterparts on ms timescales.
\end{abstract}


\section{Introduction} \label{sec:intro}

More than a decade has passed since fast-radio bursts (FRBs) were discovered \citep{Lorimer07}, and it is now well established that they are emitted by extragalactic, astrophysical sources \citep[e.g.][]{Cordes19, Zhang20}. However, the stellar systems, their configuration, and the exact physical mechanism(s) capable of releasing radio pulses required by FRB energies ($\sim 10^{40}$--$10^{43}$\,erg) and short timescales ($\sim 10^{-3}$\,s) remain elusive.

Several theories have been proposed for the origin of FRBs \citep[see][for a review]{Petroff22}, although many of these are already ruled out for the bulk of the FRB population \citep{Bhandari20a,Heintz20,Marnoch20,Gordon23}. The current prevailing view is that they may be related to eruptions from magnetars based on the detection of a low-energy FRB from the Galactic magnetar SGR\,1935+2154 \citep{Bochenek20b,CHIME20b}. However, the magnetar theory is complicated by evidence for periodicity in some FRBs \citep[][whereas magnetar eruptions are more likely to be stochastic]{CHIME20,Rajwade20} and the detection of a repeating FRB in a globular cluster with an extremely old stellar population \citep{Kirsten22}.  Curiously, the FRB signal from SGR\,1935+2154 was accompanied by a simultaneous detection of a hard X-ray emission by INTEGRAL and Konus-WIND, suggesting a broadband, non-thermal emission model \citep{Mereghetti20,Ridnaia21}.  Even in the specific case in which FRBs arise from magnetar eruptions, various models predict broadband, multi-wavelength emission via an afterglow from a synchrotron maser \citep{Waxman17,Metzger19,Margalit20}, coherent curvature radiation from charged particles in the magnetic field \citep{Kumar17,Ghisellini18,Katz18,Yang18}, or inverse Compton scattering of FRB photons to optical wavelengths \citep{Zhang22}.  A key component of most of these theories is that the optical signal will be both simultaneous with and have a similar timescale to the FRB, producing a so-called fast optical burst \citep[FOB; see, e.g.,][]{Karpov19,Yang19}.  

Unlike the heterodyne receivers that can detect FRBs as voltages sampled with variable time resolution down to microseconds or even faster \citep{Day20}, the vast majority of optical detectors operate with a fundamental limit on their exposure times of a few seconds, mostly driven by readout time and shutter speed \citep[e.g.,][]{Ivezic19}.  This presents a challenge for detecting optical counterparts to FRBs with timescales of milliseconds, although sensitive, wide-field surveys such as the Vera C. Rubin Legacy Survey of Space and Time may detect several dozen with proper filtering of their transient alerts \citep{Homar23}.

Targeted follow up of FRBs with high-speed optical cameras such as electron multiplying CCDs \citep{Alopeke} offers a better strategy for constraining optical emission with a duration of milliseconds.  By observing FRBs during periods of high activity \citep[e.g., repeating FRBs with known periods or FRBs undergoing ``burst storms'' with hundreds of events over hours or days;][]{Fonseca20,Fong21,Ravi22}, we can maximize the likelihood that an optical facility is observing a FRB when a radio burst is detected.  This strategy has been implemented by several groups for FRB\,20121102A \citep{MAGIC2018}, \frbname\ \citep{kilpatrick2021}, and FRB\,20201124A \citep{Piro21} among others, including high-speed optical camera observations of FRB\,20121102A by \citet{Hardy17} and \frbname\ by \citet{Pilia20}.  However, in both cases these observations were relatively shallow, limited by the aperture size of the telescopes used (1.2--2.4\,m) and conditions at the observing sites.

Here we present results from an observing campaign of the periodic, repeating \frbname\ with the \alopeke\ high-speed camera on the 8.1\,m Gemini-North telescope at Maunakea, Hawaii.  By targeting the FRB during expected periods of high activity and during the transit window when it was observable by CHIME, we obtained two observations simultaneous with radio bursts.  Our observing strategy, data reduction, and calibration is described in Section~\ref{sec:data}.  We describe our analysis of the data and limits on an optical counterpart to the radio bursts in Section~\ref{sec:analysis} and the implications for optical analogs, counterpart models, and prospects for future high-speed optical observations of FRBs in Section~\ref{sec:results}.  Finally, we conclude in Section~\ref{sec:conclusion}.

Throughout this paper, we assume a luminosity distance to \frbname\ of 150~Mpc from \citet{Marcote20} and \citet{CHIME20} and a Milky Way reddening to \frbname\ of $E(B-V)=0.87$~mag \citep{Schlafly11}.

%%%%%%%%%%%%%%%%%%%%%%%%%%%%%%%%%%%%%%%%%%%%%%%%%%%%%%%%%%
\section{Data and Calibration}
\label{sec:data}

\subsection{CHIME Radio Detection}\label{sec:radio}

\begin{figure}
\flushleft
\includegraphics[width=0.97\columnwidth]{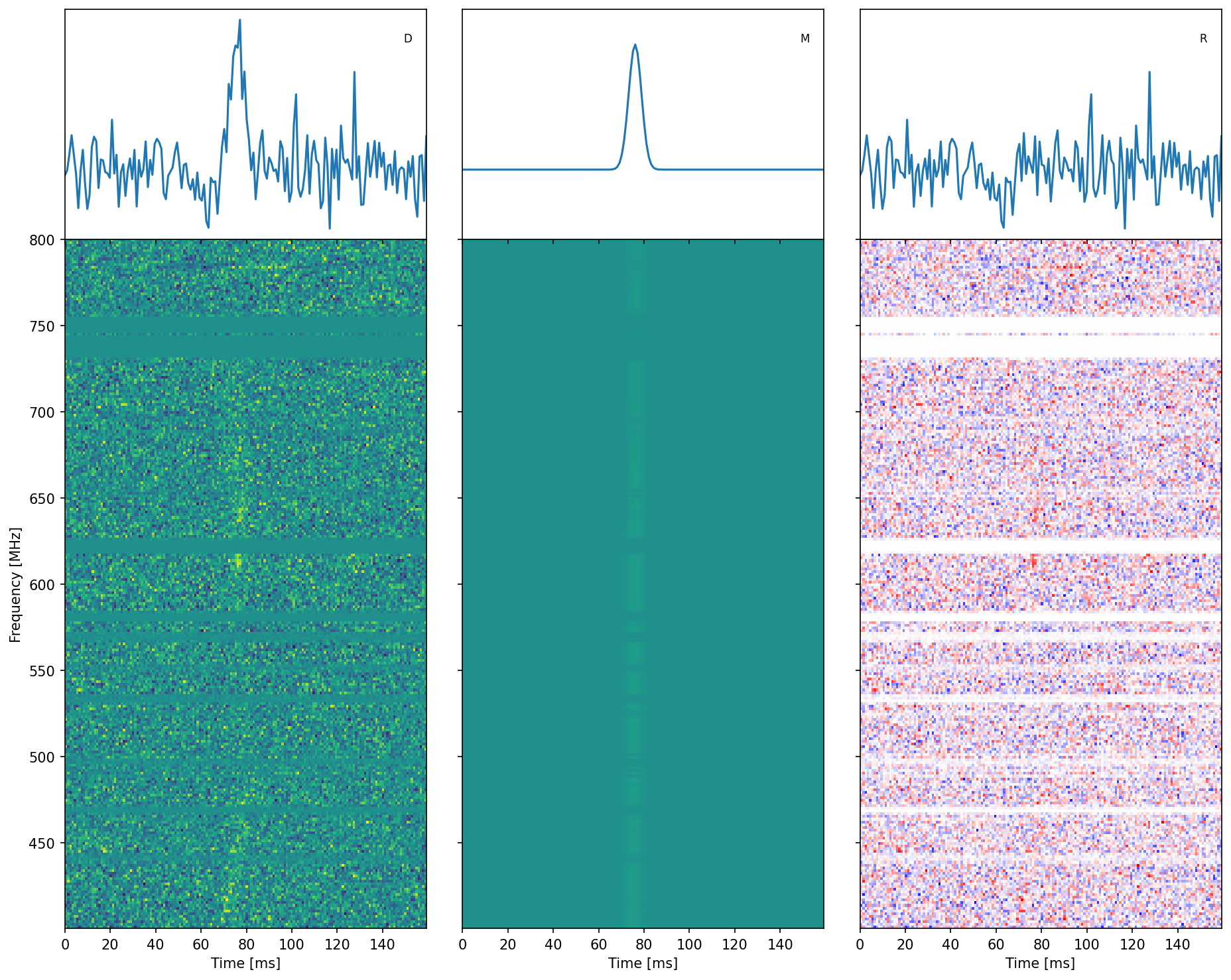}
\includegraphics[width=0.97\columnwidth]{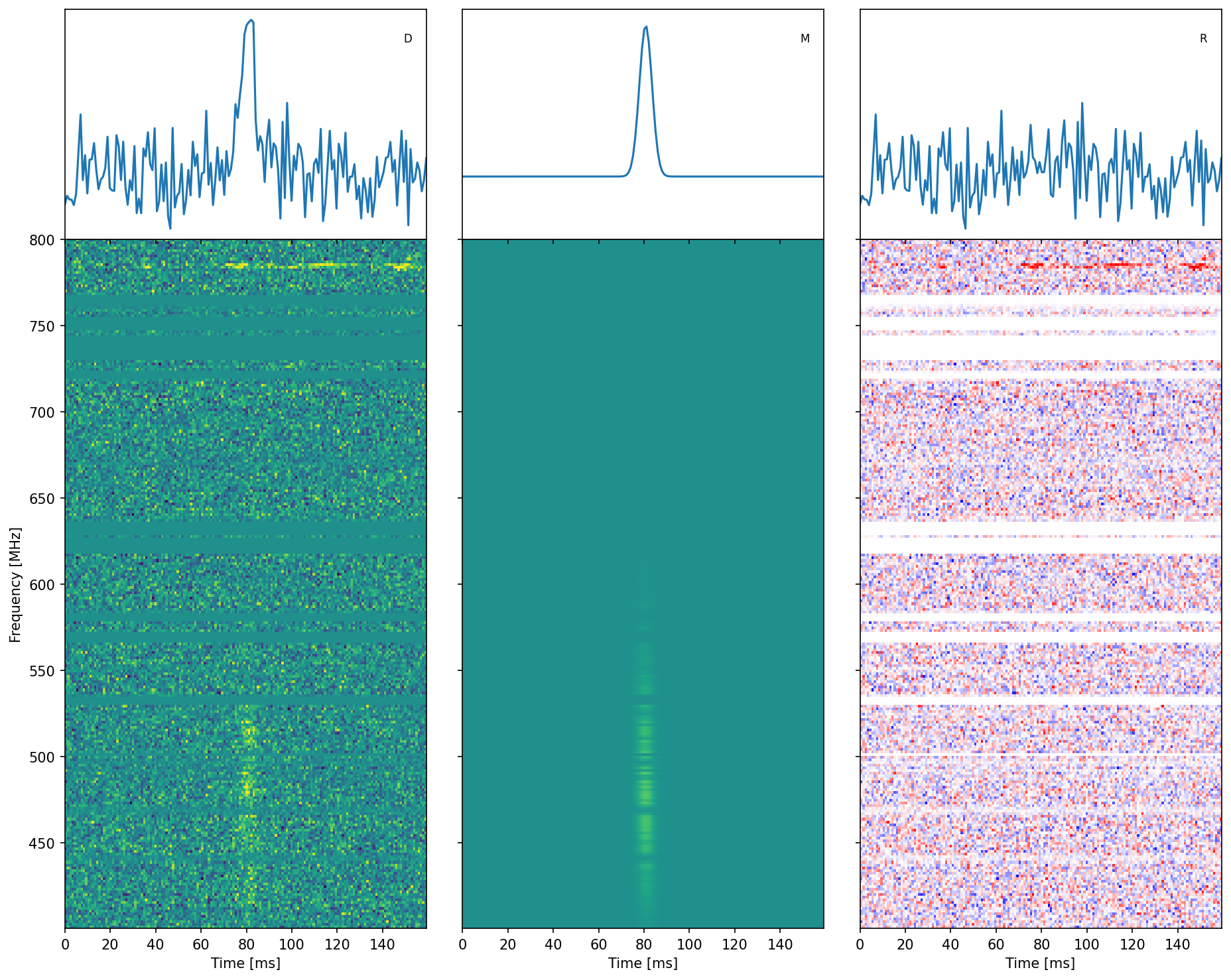}
\caption{Dynamic spectra for the CHIME detection of pulse from \frbname\ on UTC 2020-10-23 (upper panels) and 2022-09-08 (lower panels).  In each row, the figures correspond to the de-dispersed time profile in each frequency channel with the actual CHIME data as the left panel, a model fit as the middle panel, and the residual to the model fit as the right panel. The time of arrival of the pulse defines the $0$ value in the $x$-axis, which is \chimearrivala\ for the upper panels and \chimearrivalb\ for the lower panels. Bands of radio frequency interference (RFI) have been masked out for both detections.}
\label{fig:detection}
\end{figure}

CHIME detected bursts from \frbname\ as it was transiting at \chimearrivala\ and \chimearrivalb, which is confirmed by the dispersion measure (DM) from both events 350.5 and 349.9~pc~cm$^{-3}$, respectively, compared with the average DM of 349.2~pc~cm$^{-3}$ \citep[Figure~\ref{fig:detection} and][]{CHIMErep,Marcote20}.  The consistency in sky localization and DM-space rules out the possibility of chance coincidence from another burst.

The basic burst properties from \frbeventa\ and \frbeventb\ were derived using the {\tt fitburst} codebase \citep{Fonseca23}. These bursts have durations of $t_{\rm radio}=$2.7$\pm$0.3~ms and 2.7$\pm$0.2~ms, peak flux of $f_{\nu,{\rm radio}}=$0.5$\pm$0.2~Jy and 0.5$\pm$0.2~Jy, and fluence of $F_{\rm radio}=$2.6$\pm$0.8~Jy~ms and 3.5$\pm$0.8~Jy~ms, respectively.  As with all CHIME bursts \citep[e.g., those in the CHIME DR1 FRB catalog;][]{CHIMEDR1}, the dispersion-corrected arrival time is calculated at a rest frequency of 400.19~MHz, which we assume below for comparison to our optical data.

\subsection{\alopeke\ High-speed Imaging}

\begin{figure*}
\centering
\includegraphics[width=1\textwidth]{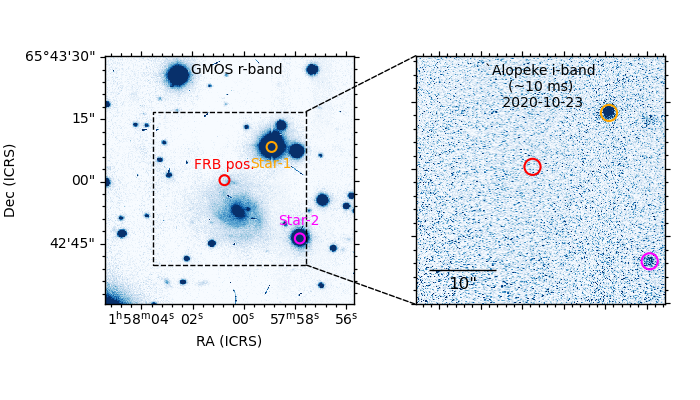}

\caption{
{\it Left}: Portion ($\sim 1 \arcmin \times 1 \arcmin$) of Gemini GMOS $z$-band image from \cite{Marcote20} centered on \frb. The FRB position (red) and nearby reference stars (\refstara; orange, \refstarb; purple) are labeled. {\it Right}: Field-of-view of \alopeke\ ($37\arcsec \times 37\arcsec$) around the \frb\ position. Only the \refstara\ and \refstarb\ are detected in a single \dtres~ms exposure.
}
\label{fig:fov}
\end{figure*}

We contemporaneously observed the \frbname\ field with the \alopeke\ high-cadence camera \citep{Alopeke,Scott21} as part of Gemini-N programs GN-2020B-DD-103 and GN-2022B-Q-202 (PI Prochaska). Observations were carried out on UTC 2020-10-23 and 2022-09-08 using \alopeke's wide-field mode with $2\times2$ binning in a region of $256 \times 256$ pixels around the FRB position. This provides an effective field-of-view of $37\arcsec \times 37\arcsec$, a pixel scale of $0.145\arcsec~{\rm pixel}^{-1}$, and a time resolution of \dtres\,ms.  

We coordinated the \alopeke\ observations to coincide with the CHIME transit at the expected peak of the $\sim$$\period$\,day periodic activity of the repeater \frbname\ \citep{CHIME20}. We observed the field for $\approx$1136\,s almost continuously in the $r$ and $i$ bands starting at UT 2020-10-23T07:41:40.265 and for $\approx$1315\,s starting at 2022-09-08T10:39:04.629. These $r$ and $i$-band exposures were observed near simultaneously using the blue and red cameras, respectively. In each camera and exposure, we obtained $5000$ individual exposures of \dtres\,ms each (with accumulation cycle time of \dttot\,ms), where each set lasted for $\sim 1$\,min including read-out overhead.  There were 20 separate exposures per camera on 2020-10-23 and 23 separate exposures per camera on 2022-09-08. 

Our observing strategy covered time-frames approximately $\approx10$--$12$ min before and after the peak of the CHIME transit on each date. As we observed on a date near the peak of the \frbname\ activity cycle, this maximized the likelihood of observing at a time when CHIME was likely to detect a radio burst.

After science observations on each night, a series of $3$ flat field calibrations were taken for each camera, with the exact same setup ($5000$ exposures per series), using the twilight sky as reference. A set of $2$ bias/dark series observations were also taken after science observations on the same day. Given the short exposure times, bias and darks are essentially the same and we combined them together to produce a ``master bias.''

We reduced all of our imaging with using custom software implementing \texttt{astropy} \citep{astropy}. Master bias, dark, and flat images were created for both cameras and filters (i.e., the blue, $r$-band and red, $i$-band, respectively) by combining the individual exposures of the individual bias frames and flat frames. We obtained an individual reduced image by subtracting the corresponding master bias to each individual frame, and by dividing the result by the corresponding normalized master flat.

The field-of-view of the observations in both epochs is centered at the FRB position, and every exposures contains at least two point-like sources that are classified as stars in the Pan-STARRS catalog \citep{Chambers16} and are bright enough to be well detected in the individual \dtres\,ms exposures. We used these two stars to define both the absolute astrometric and photometric calibration across all 5000 frames individually for every exposure.  Comparing to an overlapping, wider Gemini-N/GMOS image \citep{Marcote20}, the two stars are well aligned to this much deeper frame. We refer to the brightest star in the field-of-view \citep[Pan-STARRS objID=186860294956243663;][]{Chambers16,Flewelling16} as \refstara\ and the second brightest star (objID=186850294911316323) as \refstarb\ throughout the manuscript.

\subsection{Astrometry}\label{sec:astrometry}

For absolute astrometry, we used the positions of \refstara\ ($\alpha=29.495612$, $\delta=65.719111$) and \refstarb\ ($\alpha=29.491117$, $\delta=65.712993$) \citep{GaiaDR2, Lindegren2018} to set the alignment and rotation of the \alopeke\ camera, assuming zero distortion and an absolute pixel scale of $0.145\arcsec \, \rm pixel^{-1}$ across the entire detector.  This alignment strategy yields good results when comparing the stacked 5000 frames for each exposure to the deeper GMOS observation where we detect at least 5 point sources in the stacked frames.  We obtain $\sim$0.1\arcsec\ root-mean square offsets in both right ascension and declination between the stacked \alopeke\ data and the GMOS image. Figure~\ref{fig:fov} shows a GMOS $z$-band image  ($\sim 1\arcmin \times 1\arcmin$; left panel) and a single \alopeke\ $i$-band image ($37\arcsec \times 37\arcsec$; right panel) around the FRB position.  The seeing was approximately 0.5\arcsec\ in the first epoch and 0.8\arcsec\ in the second epoch and we perform photometry within 2~FWHM of the FRB location, so we are confident that astrometric uncertainty does not significantly affect our analysis.

\subsection{Time calibration and sensitivity}\label{sec:time}

The \alopeke\ time stamps for each of the 5000$\times$\dtshort~ms exposures are given by the Network Time Protocol (NTP) from UTC times. Its absolute time accuracy is \aata\,ms \citep[see, e.g.,][]{Alopeke,Scott21}, mostly driven by the variable lag between the computer receipt from the NTP server and the triggering of the cameras.  This sets our primary source of time calibration uncertainty. We note, however, that the relative time accuracy between individual \dtshort~ms frames in the \alopeke\ exposures is much smaller ($\sim 70$\,ns), and thus we ignore them.

CHIME operates with a time resolution of 0.983\,ms \citep{CHIME18,CHIME20}, and the uncertainty on the arrival time at infinite frequency for each burst ({\tt mjd\_inf\_err}) is typically 0.5--2~ms.  Compared with the uncertainty in the time accuracy for CHIME, we consider this to be a negligible uncertainty.

The topocentric FRB pulse time arrival at $400$\,MHz was \chimearrivala. This implies that any putative optical counterpart should have arrived $9.083$\,s earlier, that is, at \opticalarrivalCHIMEa\ based on the arrival time at infinite frequency for a radio signal detected at 400~MHz and dispersion measure of 350.19~pc~cm$^{-3}$ (Figure~\ref{fig:detection}) using equation 1 of \citet{Cordes19}.  Considering the rapid timescales involved in this calculation, we also consider the light-travel time between \alopeke\ and CHIME, which are located on Maunakea, Hawaii and Penticton, BC, Canada, respectively, separated by a direct distance of 4470~km.  This corresponds to a maximum difference in arrival times for a signal at infinity frequency of 14.9~ms between \alopeke\ and CHIME.  We targeted \frbname\ when it was transiting over CHIME, and at the arrival time at infinite frequency, it was at an hour angle 57~s east of CHIME.  This implies that the same signal would arrive approximately the full 14.9~ms light-travel time later in Hawaii, and so we assume the arrival of the signal was \opticalarrivalAlopekea\ for \alopeke.

We performed the same analysis for the 2022-09-08 burst, which was detected by CHIME at an arrival time of \chimearrivalb.  The dispersion measure of 349.8~pc~cm$^{-3}$ implies that at infinite frequency this burst arrived at the CHIME radio array at \opticalarrivalCHIMEb.  Given that the burst was at an hour angle 5~minutes and 3~seconds west of CHIME at this time and 2~hours, 18~minutes, and 21~seconds east of Maunakea, we estimate that the optical signal would arrive \alopeke\ the full 14.3~ms later at \opticalarrivalAlopekeb.  For both bursts, we take the \alopeke\ data around the corresponding arrival times calculated here to search for optical emission associated with the radio bursts, but we consider a $\pm$160~ms range of data to account for the absolute uncertainty in the \alopeke\ time stamps.

Finally, considering the cadence of $11.6$~ms and the actual individual exposure time of \dtshort~ms there is in principle the possibility that the putative optical pulse associated with the FRB arrived in between individual exposures. However, the ``down-time'' of the \alopeke\ camera in our current setup of $\sim 6\%$ is sufficiently small allowing us to be sensitive to pulses wider than 1~ms. Moreover, even for pulses intrinsically narrower than this, the fact that we are using two cameras at slightly different starting times make it very unlikely to miss a pulse in both of them.

% %%%%%%%%%%%%%%%%%%%%%%%%%%%%%%%%%%%%%%%%%%
\subsection{Flux Calibration}
\label{sec:flux}

We perform aperture photometry in the \alopeke\ data relative to \refstara\ and \refstarb. The apparent magnitudes of these reference stars are $r=\stararmag$\,mag and $i=\staraimag$\,mag for \refstara\ and $r=\starbrmag$\,mag and $i=\starbimag$\,mag for \refstarb\ obtained from \PS\ \citep{PS1}. In the following photometric analysis, we use the count rate of both stars to set the absolute flux scale for each frame.  Based on the Galactic reddening assumed above, we further correct our photometry for line-of-sight extinction of $\mardust=2.2\,{\rm mag}$ and $\maidust = 1.5$\,mag.

%%%%%%%%%%%%%%%%%%%%%%%%%%%%%%%%%%%%%%%%%%%%%%%%%%%%%%%%%

\begin{figure*}
\centering
\includegraphics[width=0.8\textwidth]{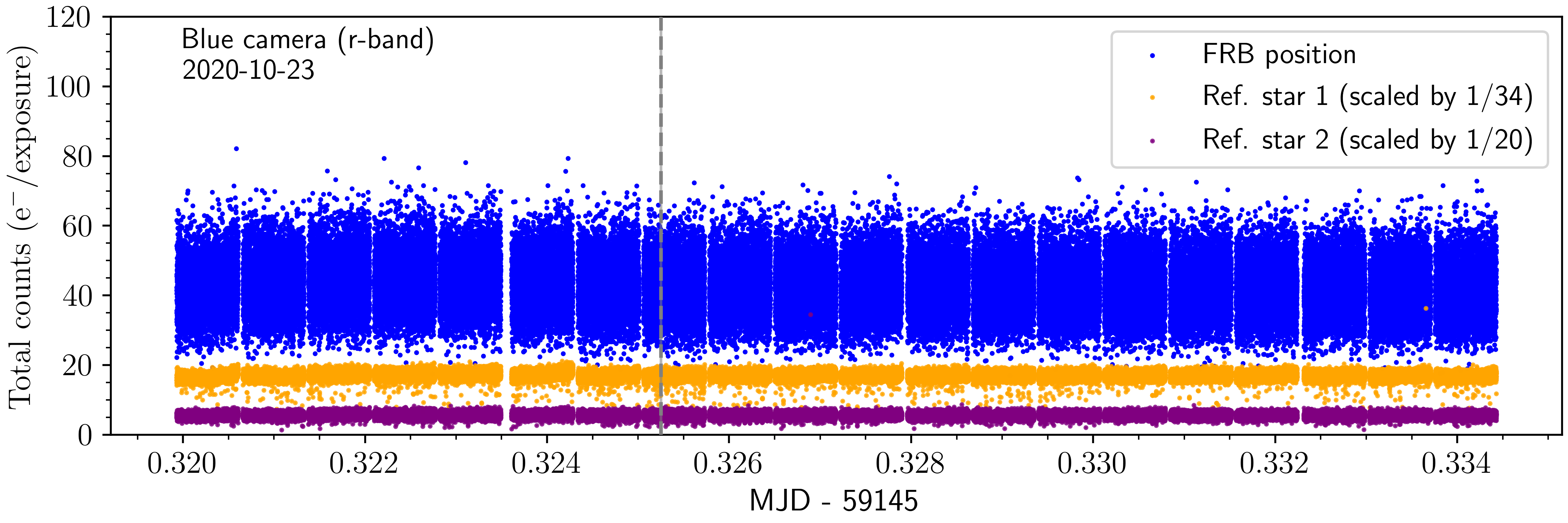}\vspace{-0.05in}
\includegraphics[width=0.8\textwidth]{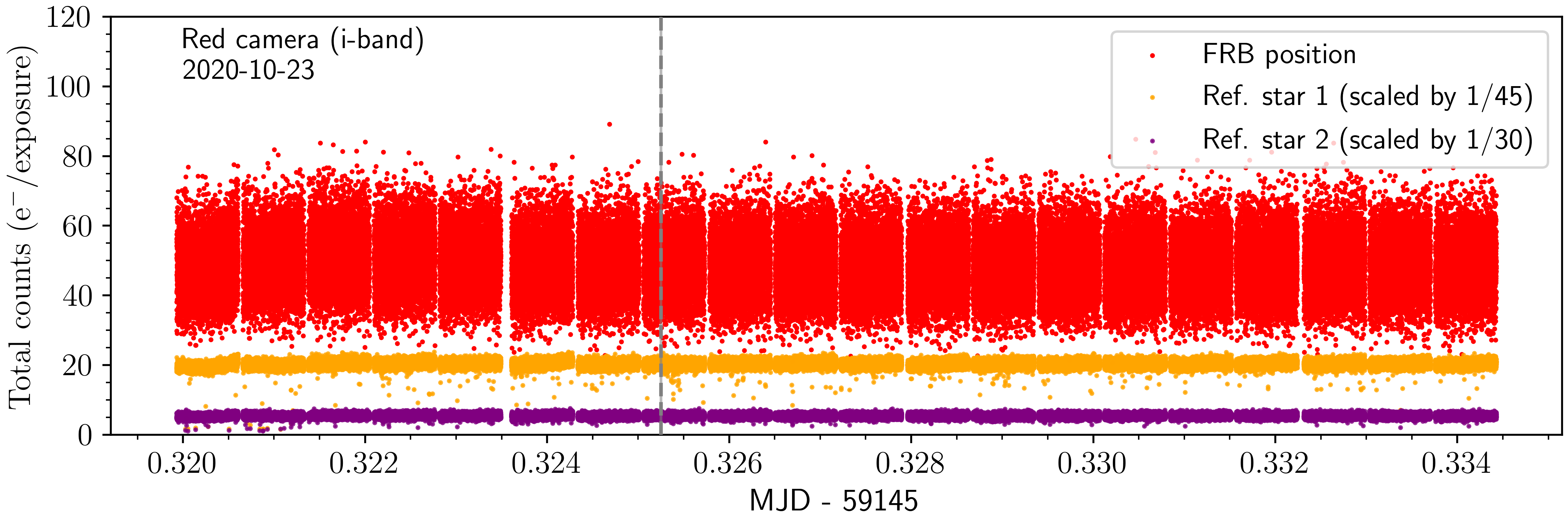}\vspace{-0.05in}
\includegraphics[width=0.8\textwidth]{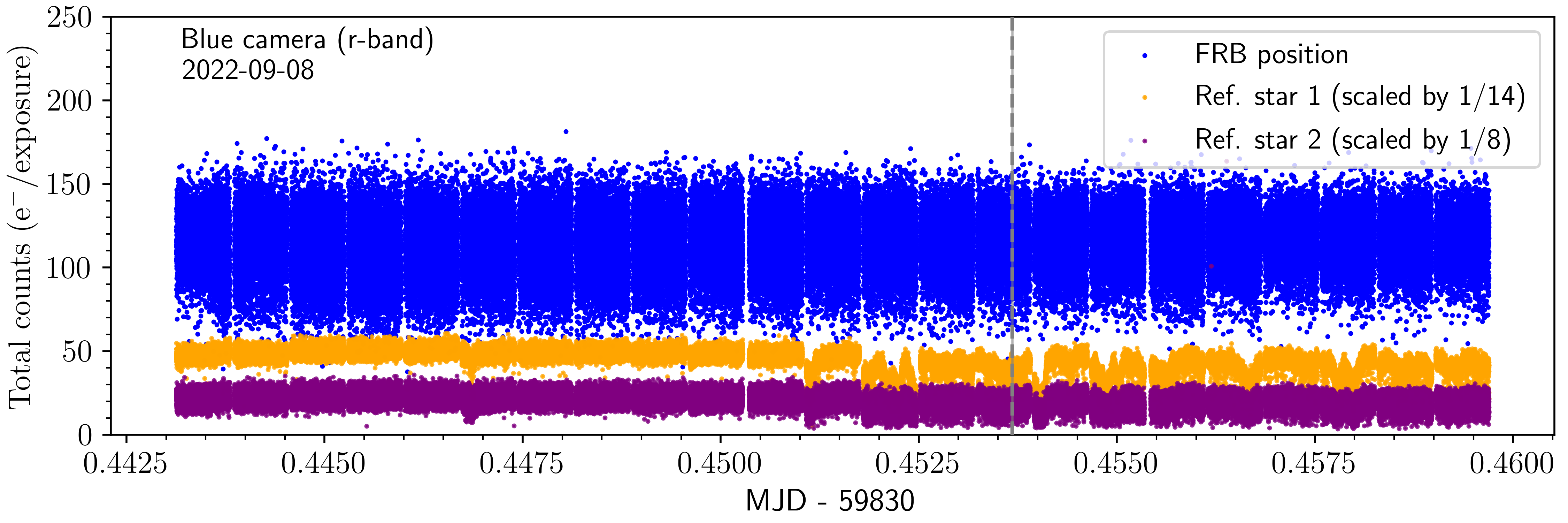}\vspace{-0.05in}
\includegraphics[width=0.8\textwidth]{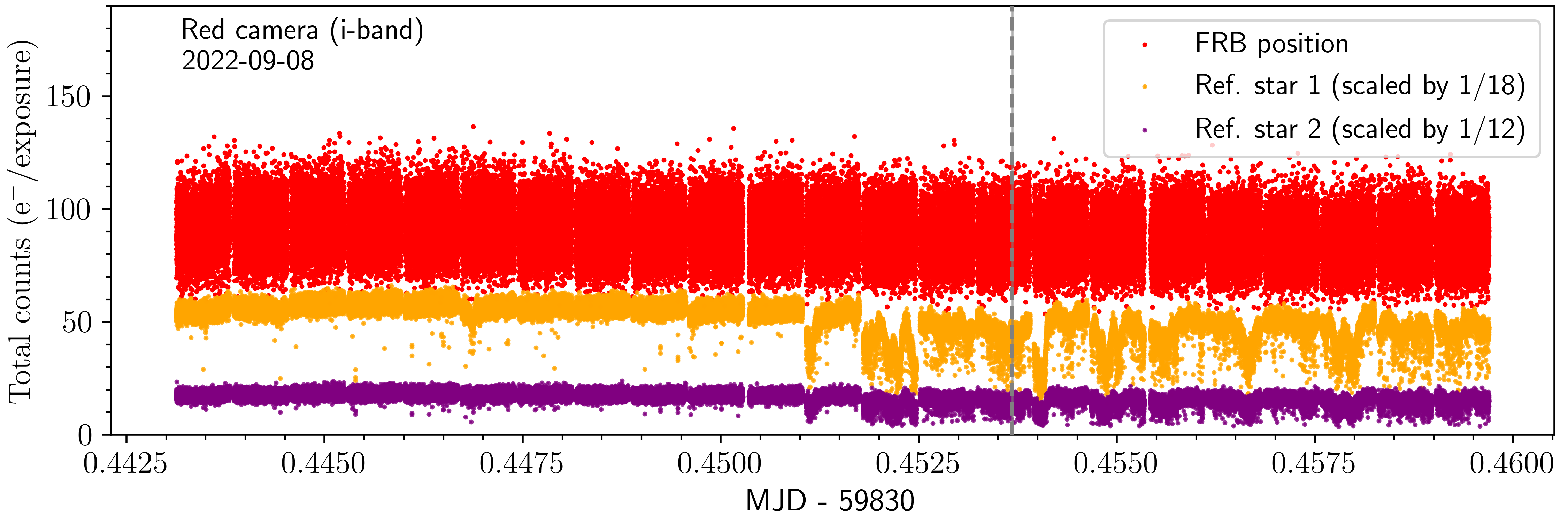}\vspace{-0.05in}
\caption{{\it Top two panels}: Total count rate (e$^-$/exposure) at the FRB location using a circular aperture with radius of two FWHM assessed from \refstara\ and \refstarb. Counts for the blue camera are shown in the top panel while those for the red camera are shown in the bottom panel. Each panel also show the respective counts of the reference star scaled by a constant factor. The count-rates show a small drift towards lower values across the observing period of $\approx 750$\,s. The regular gaps arise due to read-out of the detectors. Noticeable variation in the count level are also seen in the reference star, thus indicating they may be telluric. Any optical emission from the \frbname\ is predicted arrive at \alopeke\ at MJD = \mjdeventa\ (Section~\ref{sec:time}) marked by the dashed vertical lines. {\it Bottom two panels}: Same as the top panels for the second CHIME event we observed from \frbname.  Conditions were not photometric, as shown by the variable flux from \refstara\ and \refstarb.  The dashed line indicates the time of the burst at MJD = \mjdeventb.
}
\label{fig:counts}
\end{figure*}

\begin{deluxetable}{cccccccccccccccc}
\tablewidth{0pc}
\tablecaption{Photometry for Blue Camera ($r$-band) \label{tab:photom_blue}}
\tabletypesize{\footnotesize}
\tablehead{\colhead{MJD} 
& \colhead{\cbkg} & \colhead{\cstara} & \colhead{\cstarb} & \colhead{\ctfrb} 
} 
\startdata 
59145.32525183 & 53.95 & 629.68 & 134.18 & 53.11 \\
59145.32525169 & 33.14 & 593.92 & 119.27 & 31.15 \\
59145.32525156 & 54.34 & 593.81 & 103.40 & 40.30 \\
59145.32525142 & 45.39 & 639.92 & 107.57 & 33.92 \\
59145.32525129 & 36.45 & 637.76 & 114.95 & 42.32 \\
59145.32525115 & 31.10 & 631.27 & 127.16 & 35.86 \\
59145.32525102 & 34.27 & 600.15 & 124.73 & 51.80 \\
59145.32525088 & 45.23 & 688.93 & 118.49 & 38.60 \\
59145.32525075 & 58.60 & 222.24 & 68.41 & 51.86 \\
59145.32525062 & 59.28 & 604.63 & 111.40 & 55.51 \\
59145.32525048 & 38.98 & 652.59 & 108.91 & 36.53 \\
59145.32525035 & 49.33 & 649.81 & 124.12 & 45.42 \\
59145.32525021 & 38.35 & 581.59 & 112.30 & 54.41 \\
59145.32525008 & 45.60 & 615.87 & 117.90 & 44.30 \\
59830.45367901 & 117.72 & 722.42 & 208.14 & 94.63 \\
59830.45367888 & 141.90 & 775.88 & 251.05 & 117.41 \\
59830.45367874 & 101.42 & 718.38 & 161.78 & 132.69 \\
59830.45367861 & 134.09 & 675.11 & 154.49 & 90.93 \\
59830.45367848 & 96.41 & 687.47 & 164.43 & 133.75 \\
59830.45367834 & 118.88 & 697.87 & 191.42 & 130.06 \\
59830.45367821 & 126.88 & 714.53 & 158.47 & 127.82 \\
59830.45367807 & 92.48 & 654.53 & 173.02 & 124.73 \\
59830.45367794 & 103.90 & 686.17 & 169.42 & 133.45 \\
59830.45367781 & 131.40 & 703.54 & 184.31 & 93.70 \\
59830.45367767 & 144.47 & 690.74 & 169.79 & 119.77 \\
59830.45367754 & 89.00 & 696.97 & 149.41 & 122.43 \\
59830.45367740 & 102.38 & 765.27 & 161.21 & 107.02 \\
59830.45367727 & 77.70 & 668.95 & 160.54 & 130.89 \\
\hline 
\enddata 
\end{deluxetable} 

\begin{deluxetable}{cccccccccccccccc}
\tablewidth{0pc}
\tablecaption{Photometry for Red Camera ($i$-band) \label{tab:photom_red}}
\tabletypesize{\footnotesize}
\tablehead{\colhead{MJD} 
& \colhead{\cbkg} & \colhead{\cstara} & \colhead{\cstarb} & \colhead{\ctfrb} 
}  
\startdata 
59145.32525190 & 56.38 & 968.71 & 162.36 & 39.78 \\
59145.32525176 & 48.62 & 943.33 & 157.73 & 36.48 \\
59145.32525163 & 44.47 & 937.69 & 181.28 & 53.24 \\
59145.32525149 & 52.45 & 986.96 & 159.19 & 43.93 \\
59145.32525136 & 46.44 & 982.17 & 153.60 & 38.64 \\
59145.32525122 & 45.74 & 921.25 & 161.28 & 50.57 \\
59145.32525109 & 46.11 & 1011.73 & 178.21 & 45.82 \\
59145.32525095 & 49.06 & 953.44 & 183.01 & 72.33 \\
59145.32525082 & 56.08 & 934.66 & 173.28 & 44.09 \\
59145.32525069 & 46.95 & 939.67 & 174.52 & 45.19 \\
59145.32525055 & 71.07 & 983.18 & 168.23 & 42.54 \\
59145.32525042 & 62.22 & 1042.15 & 193.04 & 55.64 \\
59145.32525028 & 52.06 & 926.46 & 144.29 & 55.04 \\
59145.32525015 & 49.63 & 939.32 & 160.76 & 34.23 \\
59830.45367898 & 102.55 & 1026.24 & 254.88 & 97.34 \\
59830.45367885 & 96.79 & 1031.09 & 227.96 & 73.23 \\
59830.45367872 & 88.02 & 1018.30 & 204.55 & 98.24 \\
59830.45367858 & 81.72 & 1020.87 & 218.72 & 94.50 \\
59830.45367845 & 86.83 & 1089.42 & 250.15 & 96.88 \\
59830.45367831 & 85.25 & 1034.26 & 207.45 & 76.77 \\
59830.45367818 & 85.08 & 1093.24 & 240.30 & 95.13 \\
59830.45367804 & 88.20 & 1086.52 & 205.34 & 87.05 \\
59830.45367791 & 102.85 & 1065.19 & 224.98 & 90.81 \\
59830.45367777 & 100.17 & 1064.64 & 225.16 & 94.77 \\
59830.45367764 & 96.77 & 995.53 & 233.08 & 96.39 \\
59830.45367750 & 91.61 & 1047.45 & 208.60 & 94.72 \\
59830.45367737 & 66.93 & 1053.33 & 199.60 & 76.86 \\
59830.45367724 & 92.91 & 1007.54 & 261.67 & 76.40 \\
\hline 
\enddata 
\end{deluxetable}

%%%%%%%%%%%%%%%%%%%%%%%%%%%%%%%%%%%%%%%%%%%%%%%%%%%%%%%%%
%%%%%%%%%%%%%%%%%%%%%%%%%%%%%%%%%%%%%%%%%%%%%%%%%%%%%%%%%

\section{Photometric Analysis}\label{sec:analysis}

In this section we describe the analysis related to photometric measurements from the \alopeke\ imaging, including our measurements of the count rate and upper limits on the count rate at the site of \frbname\ within each $\sim$\dtshort~ms frame.

\subsection{Count-rate Measurements}

In each camera and for each exposure we first measured the centroid and full-width at half-maximum (FWHM) for reference \refstara\ and \refstarb.  We then measured the count rates of these stars within an aperture of diameter 2~FWHM and then the counts in the same sized aperture at the location of \frbname.

Our final count-rate measurements are tabulated in Table~\ref{tab:photom_blue} and \ref{tab:photom_red} for the closest $\approx$163~ms to the arrival time of \frbeventa\ and \frbeventb. These counts include all sources of photoelectrons: the night sky, the galaxy hosting \frbname, the detector, and the individual sources of interest. We refer to count-rates at the \refstara, \refstarb, and at the location of \frbname\ as \cstara, \cstarb, and \ctfrb, respectively. Our final results will be derived from departures (or lack thereof) from the mean of \ctfrb.  For comparison, we also estimate the local background count-rate near \refstara\ by measuring the counts per frame and pixel in an annulus with inner radius 3~FWHM and outer radius 6~FWHM and rescaling the total count rate to the size of the 2~FWHM.

\subsection{Constraints on a Counterpart to \texorpdfstring{\frbname}\, and Flux Limits}\label{sec:limits}

\begin{figure}
\centering
\includegraphics[width=0.49\textwidth]{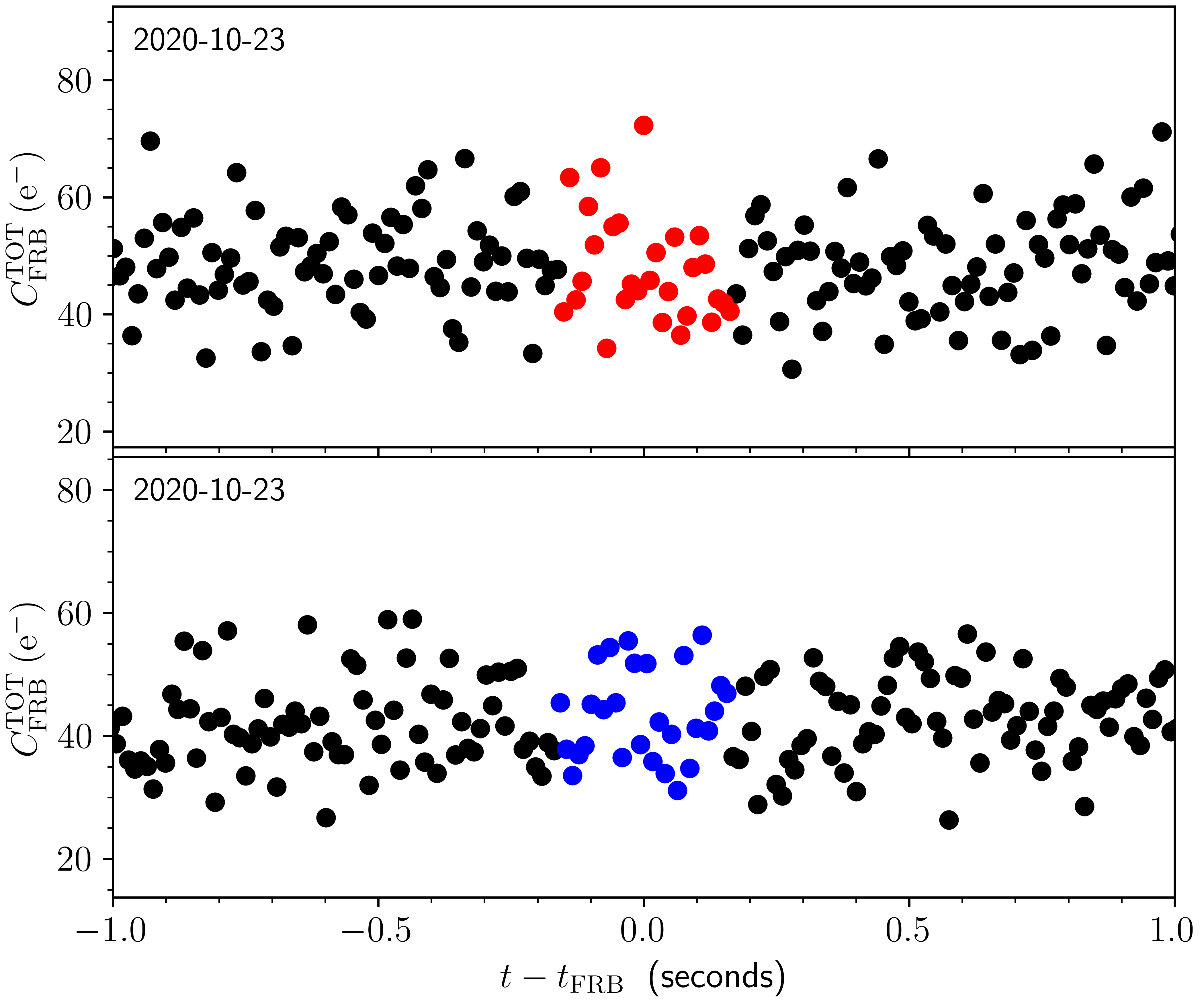}
\includegraphics[width=0.49\textwidth]{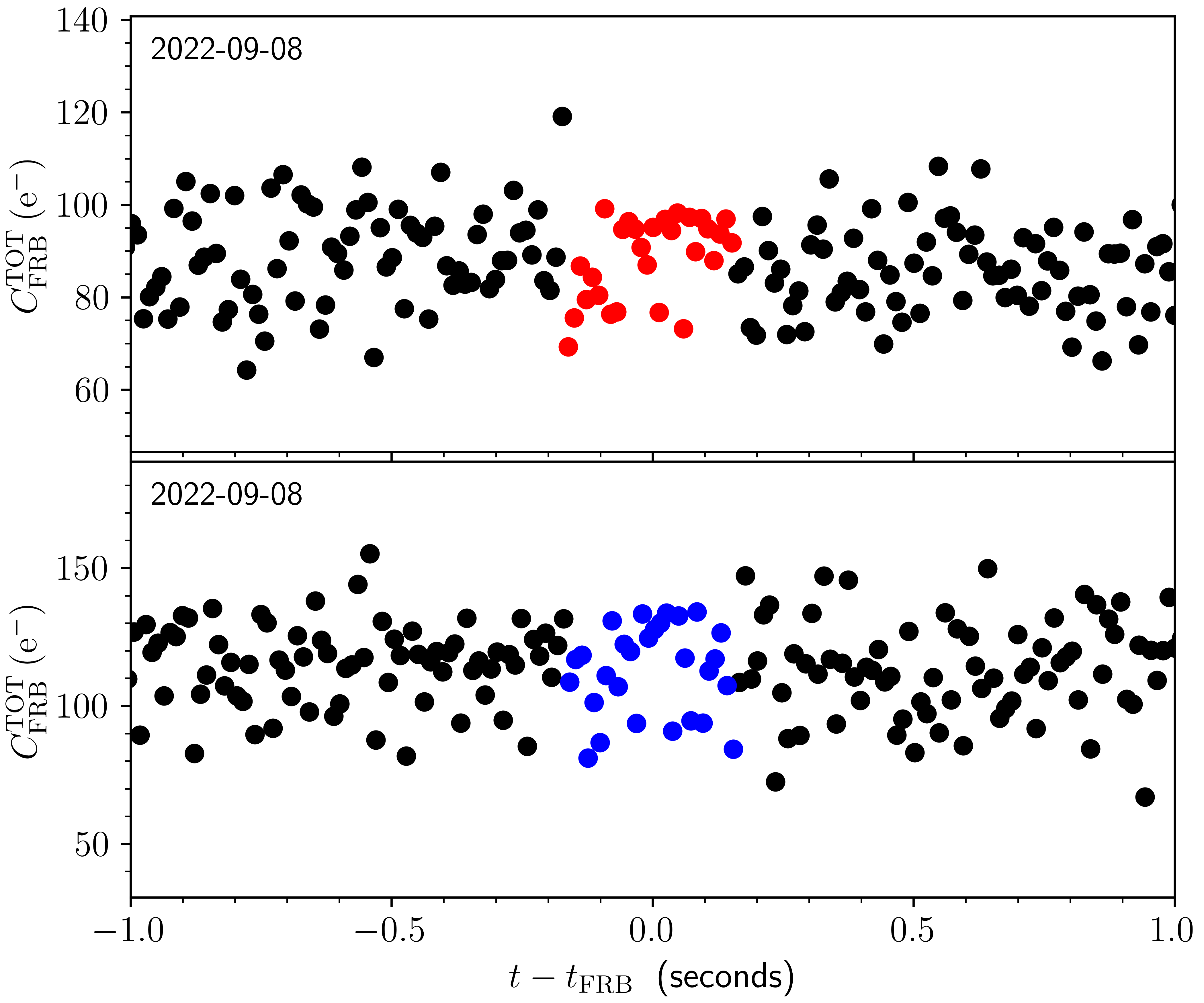}
\caption{Total counts per exposure in each camera at the FRB location for the $\approx 2$\,s around the expected arrival time of optical emission for \frbeventa\ (upper two panels) and \frbeventb\ (lower two panels). The colored dots show the $\Delta t \approx \pm 163$\,ms interval corresponding to the absolute timing uncertainty of the \alopeke\ cameras. There is no evident excess emission for any exposure during these intervals.
}\label{fig:FRB_zoom}
\end{figure}

Figure~\ref{fig:FRB_zoom} shows the subset of count measurements near the predicted arrival time for the optical emission of \frbeventa\ and \frbeventb. Within the time interval corresponding to the systematic uncertainty of the absolute timing for the \alopeke\ camera around \frbeventa\ (i.e., $\pm$162~ms from the time calculated in Section~\ref{sec:time}), the red camera recorded 28~measurements with a maximum of $\mmxfrb = 57.8$\,\cunit\ or less than $2\sigma$ from the mean count-rate during the full set of observations. Accounting for the multiple measurements within the time interval, the percentage of random draws with one or more measurements having $\mcfrb > \mmxfrb$ is 81\%. Results for the blue camera are similar with $\mmxfrb = 58.9$\,\cunit.  We repeat this analysis for \frbeventb, finding $\mmxfrb=99.2$\,\cunit, 1.2$\sigma$ above the mean in the red camera, and $\mmxfrb=134.2$\,\cunit, 1.3$\sigma$ above the mean in the blue camera.  We conclude that any prompt optical emission associated with either radio burst is not detected.  Furthermore, we report that we do not see any sources of emission at $>$10$\sigma$ at the site of \frbname\ across any of our data sets.

We proceed to estimate a conservative upper limit to the optical fluence of the FRB in both epochs and cameras.  We generate Monte Carlo realizations of the experiment by generating mock observed counts at the FRB location during the event:

\begin{equation}
\ctfrb = \mcofrb + \mcfrb
\end{equation}

\noindent with \cofrb\ described by the probability density function (PDF) of counts taken off the event (i.e., background) and \cfrb\ is drawn from a PDF for FRB emission in a single \dtshort~ms frame. For the former, we simply adopt the encircled flux throughout each entire observation (i.e., the values shown in Figure~\ref{fig:counts}), which is relatively constant throughout both data sets. For the latter, we assume a Poisson PDF with mean \mufrb\ and that the emission is limited to a single exposure.  We draw 100 realizations of the $\approx 100,000$ $\mcfrb$ measurements, increment these by random draws from the Poisson PDF for the FRB, and assess the fraction that exceed \mxfrb. 

\begin{figure}
\centering
\includegraphics[width=0.49\textwidth]{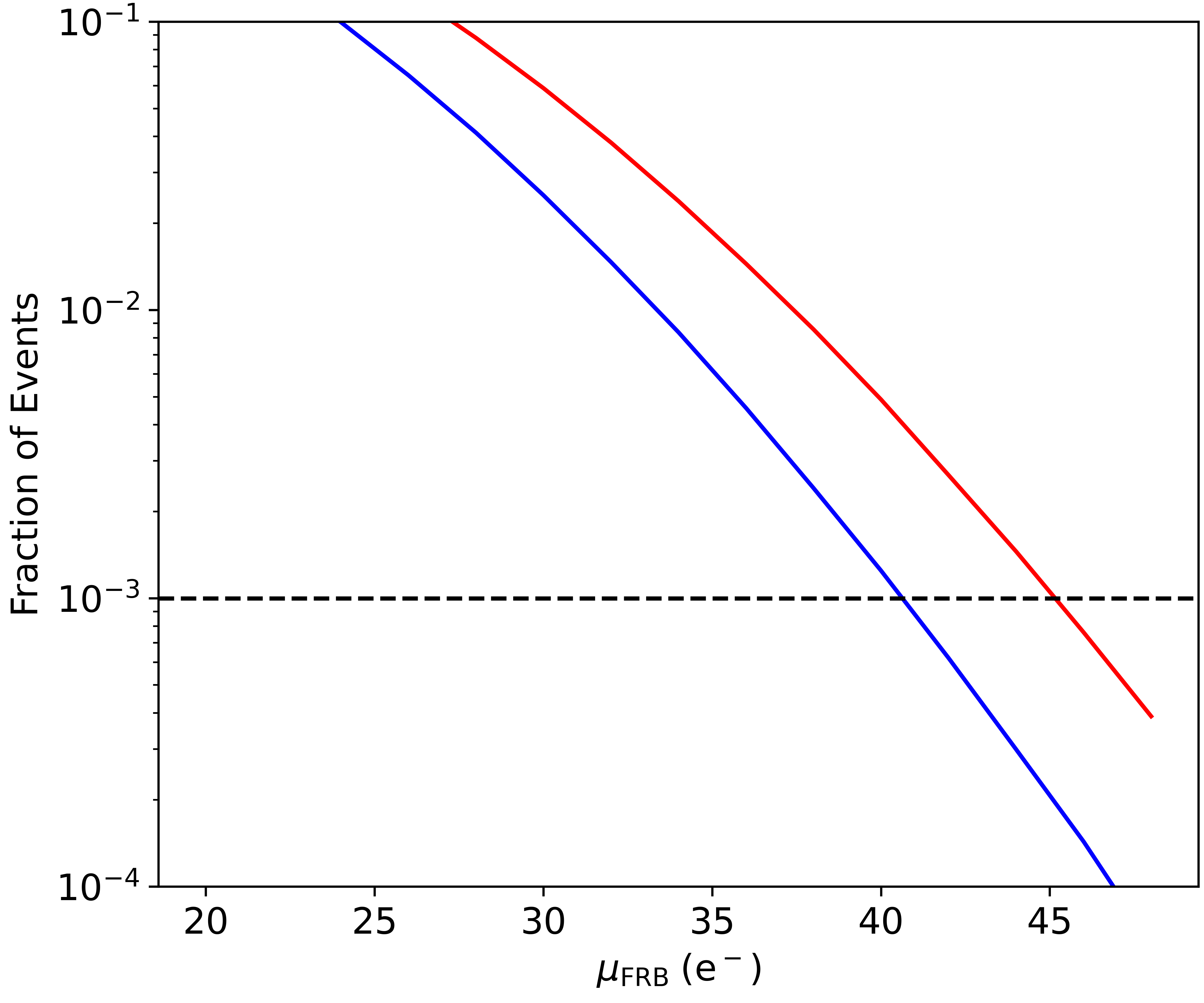}
\caption{Fraction of simulated events that would not exceed the observed maximum counts \mxfrb\ in the FRB arrival interval  as a function of assumed mean counts \mufrb\ (Poisson) contributed to a single exposure by the FRB emission. Adopting a confidence level of 99.9\%\ (corresponding to the horizontal dashed line), we set an upper limit of  $\mumufrb =  38.9\, \rm e^-$ (blue line) and $\mumufrb = 51.6 \, \rm e^-$ for the red camera (red line) for \frbeventa\ and $\mumufrb =  76.2\, \rm e^-$ (blue line) and $\mumufrb = 42.0 \, \rm e^-$ for the red camera (red line) for \frbeventb.
}
\label{fig:FRB_upper}
\end{figure}

Figure~\ref{fig:FRB_upper} shows the results for a range of \mufrb.  For \frbeventa, we find that 99.73\%\ of a random ensemble would exceed \mxfrb\ for (blue camera) $\mumufrb \approx 38.9$\,\cunit\ and (red camera) $\mumufrb \approx 51.6$\,\cunit.  Similarly, we find that for \frbeventb, 99.73\% of the ensemble would exceed \mxfrb\ for (blue camera) $\mumufrb \approx 76.2$\,\cunit\ and (red camera) $\mumufrb \approx 42.0$\,\cunit. In the following, we use these single-exposure count-rate upper limits for constraining the FRB optical emission.

\subsection{Time-variable Sensitivity Function for \alopeke}

Key to our optical fluence limits are the photometric accuracy with which we can measure the count rate from \refstara\ and \refstarb\ in each image frame. Figure~\ref{fig:counts} shows the measured counts in each camera at the FRB location \ctfrb, \refstara, and \refstarb\ for the full duration of all exposures.  In the first radio burst \frbeventa, the \alopeke\ data around the optical arrival time show a small gradient in the count-rate for the individual exposures of \refstara\ in time that we measure from a linear fit to be $d\cstara/dt \approx 0.043$ and $\approx 0.015 \; {\rm e^- \, s^{-1}}$ for the blue and red cameras, respectively, in a 2~minute window around the time of the radio burst.  This drift is a small fraction of the average count rate for \refstara\ in both cameras during this time interval (see Table~\ref{tab:photom_blue} and Table~\ref{tab:photom_red}), implying that the sensitivity function can be approximated from the entire ensemble of data with a source having a flux 1\,\cunit\ corresponding to 22.35~AB~mag in the red camera and 22.43~AB~mag in the blue camera.

\begin{figure}
\centering
\includegraphics[width=0.49\textwidth]{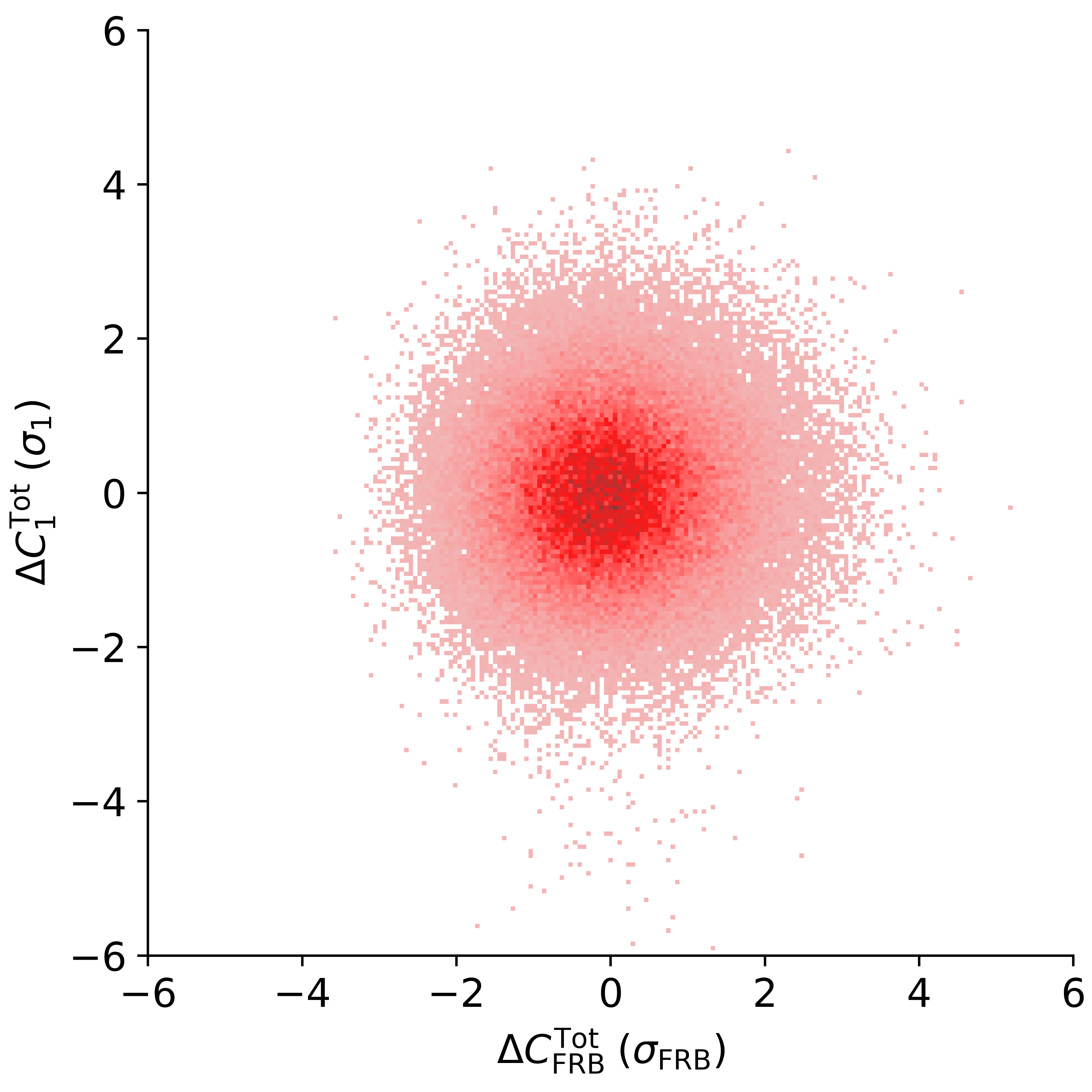}
\includegraphics[width=0.49\textwidth]{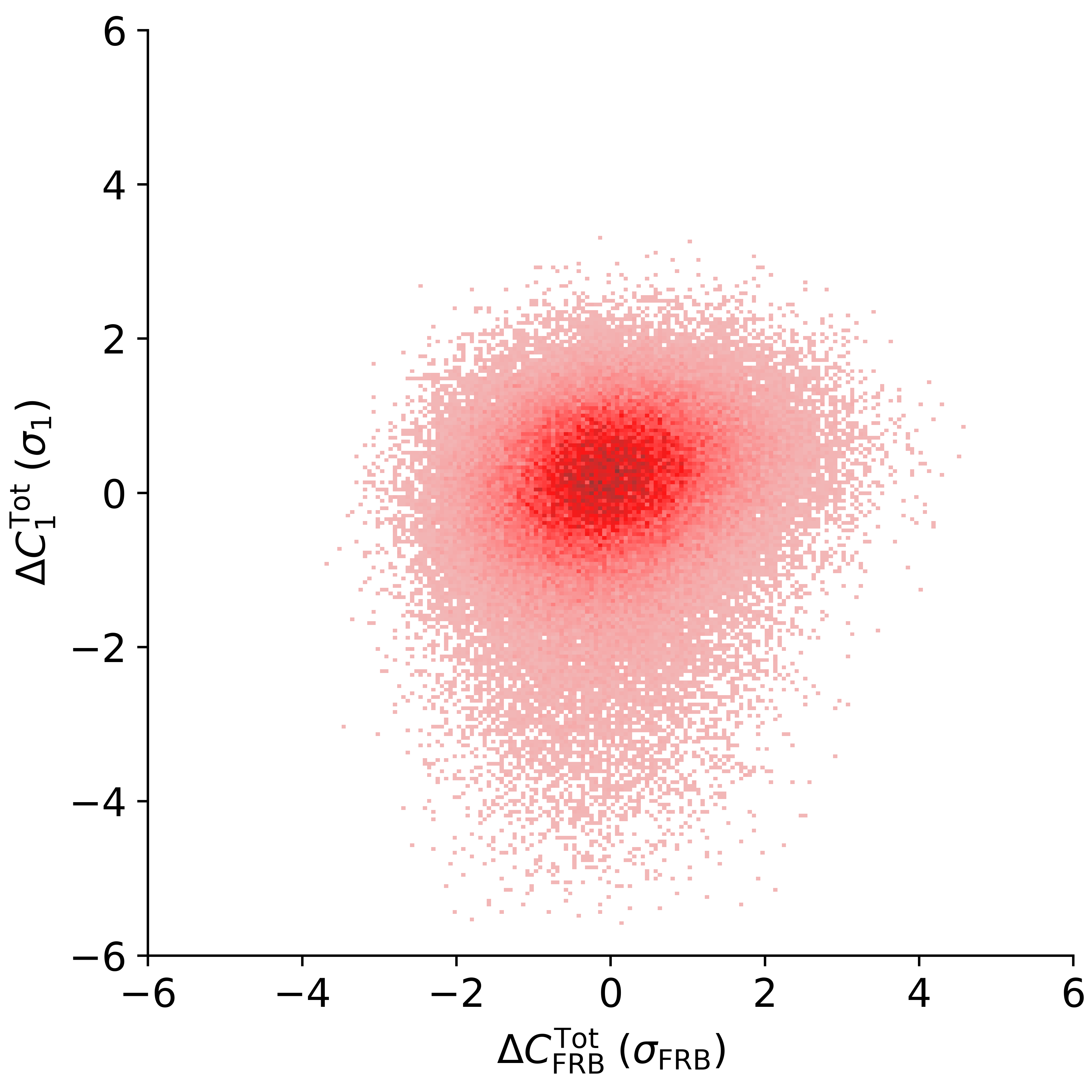}
\caption{{\it Top panel}: Deviations from the mean in units of standard deviation for the \alopeke\ measurements at the \refstara\ against those at the FRB location during the \frbeventa.  The nearly symmetric distribution indicates that the variations in count-rate are dominated by statistical fluctuations.  {\it Bottom panel}: Same as the top panel but for the \alopeke\ observations obtained during \frbeventb.  Significant variation from the mean toward negative values for \refstara\ indicate that many of these observations were obtained in non-photometric conditions, also seen in Figure~\ref{fig:counts}.
}
\label{fig:corr}
\end{figure}

However, as shown in the overall count rates of \refstara\ and \refstarb\ from the second burst on 2022-09-08, the flux from \refstara\ and \refstarb\ is significantly variable, especially across the second half of the observation when the radio burst occurred.  This effect is correlated across the blue and red detectors and both stars, indicating that it is most likely due to grey opacity due to clouds and thus it affects both detectors simultaneously.  We further demonstrate this effect in Figure~\ref{fig:corr}, which presents the individual photometric measurements at the FRB location versus those of  \refstara\ in units of standard deviation off the mean. The nearly symmetrical distribution during the first observation epoch indicates that the observed fluctuations are uncorrelated, that is the fluctuations in the counts are dominated by random, statistical fluctuations as opposed to systematics (e.g., clouds).  However, for the second epoch, there is significant variation toward negative residuals in the flux from \refstara, indicating that the star is frequently obscured by opacity in the atmosphere throughout each exposure.

We do not expect these effects to vary significantly within the 163~ms range we consider to derive the count-rate upper limit, therefore we estimate the encircled flux at the FRB position (i.e., in units of $\mu$Jy) by deriving the zero point for every frame from \cstara\ and \cstarb\ compared with their Pan-STARRS $r$ and $i$-band magnitudes.  The count rates in Table~\ref{tab:photom_blue} and Table~\ref{tab:photom_red} demonstrate that in both epochs, the count rate varied within expectations for a Poisson distribution.  Indeed, for both \refstara\ and \refstarb, the flux appears higher in the second epoch, implying that the throughput for the Gemini-N/\alopeke\ system was higher at that time and our limits are more constraining in spite of fluctuations in the atmospheric transmission.

Taking the average zero point derived jointly from both stars within the $\pm$163~ms window around each radio burst arrival time, we find that the 99.73\% confidence interval count-rate limits in Section~\ref{sec:limits} correspond to $F_{r} < 1.38~\mu$Jy~s and $F_{i} < 2.01~\mu$Jy~s for \frbeventa\ and $F_{r} < 3.27~\mu$Jy~s and $F_{i} < 1.84~\mu$Jy~s for \frbeventb\ before correcting for Milky Way dust and within each 10.4~ms observation.  For reference, these limits correspond to a magnitude limit of $m_{r} > 18.6$~AB~mag and $m_{i} > 18.2$~AB~mag for \frbeventa\ and $m_{r} > 17.7$~AB~mag and $m_{i} > 18.3$~AB~mag for \frbeventb\ also before correcting for Milky Way dust and within each 10.4~ms observation.

After we correct for line-of-sight extinction and at the assumed distance to \frbname, we estimate the isotropic equivalent specific energy within each band of $E_{\nu} = 4 \pi D_{L}^{2} f_{\nu}$.  Along with the effective frequency of each waveband assuming $\lambda_{\rm eff}$ of 6231 and 7625~\AA\ for $r$- and $i$-bands, respectively, we derive $\nu E_{\nu, r} < 1.4\times10^{41}$~erg and $\nu E_{\nu, i} < 8.8\times10^{40}$~erg for \frbeventa\ and $\nu E_{\nu, r}<3.2\times10^{41}$~erg and $\nu E_{\nu, i}<8.1\times10^{40}$~erg for \frbeventb.  We adopt these values for comparing to the radio fluence of each burst and the multi-wavelength energetics of \frbname\ in the following discussion.

\section{Results and Discussion}\label{sec:results}

\begin{figure}
    \includegraphics[width=0.49\textwidth]{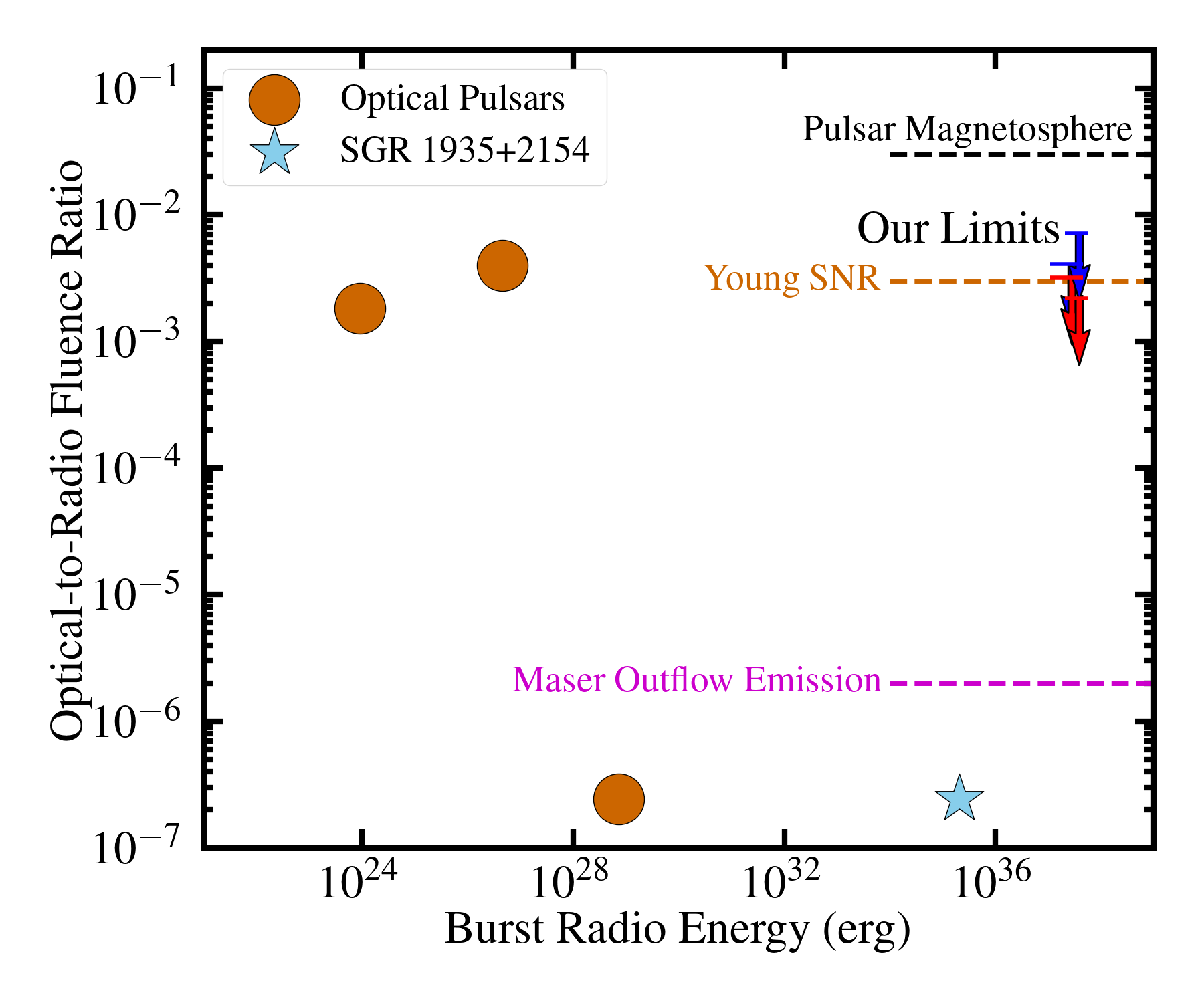}
    \caption{Our limits (downward arrows) on the optical-to-radio fluence ratio from \frbname\ in $r$- (blue) and $i$-bands (red).  We also compare to the optical-to-radio fluence ratios from the Crab, Geminga, and Vela optical pulsars (Section~\ref{sec:pulsars}) as well as the expected ratio for the bursts from the Galactic magnetar SGR\,1935+2154 (Section~\ref{sec:magnetar}). 
 Finally, we also plot the approximate fluence ratios (dashed lines) predicted for various FRB progenitor and emission models from \citet{Yang19}, including inverse Compton emission from a pulsar magnetosphere or young supernova remnant (SNR) as well as synchrotron maser emission from an outflow from a magnetar.}\label{fig:eta}
\end{figure}

Compared with the radio fluence of \frbeventa\ and \frbeventb\ (Section~\ref{sec:radio}), our limits correspond to optical-to-radio fluence ratios of $\eta_{\nu} \equiv F_{\rm opt}/F_{\rm radio}<$2--7$\times10^{-3}$ \citep[note that similar to, e.g.,][we use the $\nu$ subscript to distinguish from the ratio of the total energy radiated in each band]{Chen20}. Figure~\ref{fig:eta} shows our fluence ratio limits versus the radio energy and are comparable to the upper end of fluence ratios from optical pulsars, the expected broadband optical counterpart from SGR\,1935+2154 \citep[using analysis in][as described below]{De20}, and various progenitor models presented in \citet{Yang19}. These are the deepest limits to date for any radio burst on timescales $\lesssim$10~ms, providing useful constraints for the progenitor system and emission model powering the rapid but energetic radio burst. Throughout this section, we analyze these limits in the context of predictions for fast optical burst counterparts to FRBs.

\subsection{Constraints on the Optical Energetics}\label{sec:energetics}

Compared with previous efforts to observe optical emission from FRB\,20121102A using ULTRACAM \citep{Hardy17}, our fluence limits are a factor of 100$\times$ more constraining in the same bandpass and over timescales a factor of $\approx$7$\times$ faster (\dtshort~ms versus 70.7~ms).  Given that \frbname\ is $>$6$\times$ closer than FRB\,20121102A \citep[which is at a redshift $z=0.193$ or $D_{L}=950$~Mpc;][]{Tendulkar17} and has a comparable line-of-sight extinction, our constraints on the energy scale of any optical burst are therefore $\approx$4500$\times$ more constraining.  Similar efforts targeting millisecond-timescale optical emission from FRBs have been conducted with the Tomo-e Gozen high-speed CMOS camera observing 11 bursts of FRB\,20190520B \citep{Niino20} and the photomultiplier SiFAP2 and fast optical cameras Aqueye+ and IFI/Iqueye+ targeting \frbname\ \citep{Pilia20}.  These observations resulted in energy limits in a wide passband $T$ (370--730~nm) of $\nu E_{\nu, T}=2.9\times10^{43}$~erg on FRB\,20190520B on a timescale of 40.9~ms and in $V$-band on $\nu E_{\nu, V}=1.5\times10^{41}$~erg on a timescale of 1~ms for \frbname.  Our limits on energy are significantly more constraining on similar timescales, yielding the best constraints to date on millisecond timescale optical emission contemporaneous with a FRB.

\subsection{Comparison to Optical Emission from Pulsars}\label{sec:pulsars}

Pulsars observed in the Milky Way galaxy are among the closest analogs to extragalactic FRBs that also have optical detections simultaneous with their radio bursts, with seven such known ``optical pulsars'' \citep[][]{Cocke69,Peterson78,Middleditch87,Shearer97,Shearer98,Kern03,Slowikowska09,Ambrosino17}.  The best-studied example is the Crab pulsar \citep[see][for a review]{Buhler13}, which exhibits radio pulses known to correlate with enhanced optical pulse emission \citep{Shearer03}, a progenitor model that has been extrapolated up to higher burst energies for some FRBs \citep{Lyutikov16}.  The optical pulses are characteristically wider in time than the radio pulses roughly by a factor of 5, with the peak of the emission arriving before that of the radio pulse.

These pulses can exhibit a range of optical fluence ratios, but on average are measured to have $\eta_{\nu}\approx10^{-3}$ \citep[][see their phase-averaged emission in Fig.~2]{Buhler13}. We show them for comparison in Figure~\ref{fig:eta} as orange circles. Note that this quantity depends sensitively on the choice of radio band used in normalizing the fluence ratio due to the steep spectral indices of Crab pulses from $\nu^{-2.2}$ to $\nu^{-4.9}$ \citep{Karuppusamy10}.  Here we choose the spectrum of the Crab pulsar at 400~MHz from \citet{Buhler13} for direct comparison to the CHIME radio fluence from \frbname\ and assume an average radio burst duration of 300~$\mu$s \citep[consistent with][]{Shearer03} in deriving the emitted radio energy per pulse.  We also compared to the emitted radio energy from the Geminga and Vela pulsars (also shown as orange circles in Figure~\ref{fig:eta}), whose phase-averaged spectral energy distributions are presented in \citet{Danilenko11} for Geminga and \citet{Mignani17} for Vela.  For the former, we assume a 400~MHz brightness of 100~$\mu$Jy, and exhibit a large range in optical-to-radio fluence ratios from $\sim2\times10^{-3}$ to $2\times10^{-7}$ for Geminga and Vela, respectively.

While it is informative to investigate the ratio of radiated optical energy on short timescales for bursts from other neutron stars, we note that our limits require a moderately lower optical energy than the Crab and Geminga pulsars.  It will be challenging to rule out optical bursts with total fluences four orders of magnitude less energetic than our limits and similar to Vela without a burst closer than a few Mpc whose emitted radio energy is comparable to \frbname.  We therefore turn to other sources and emission models more directly comparable to \frbname\ and theoretically capable of partitioning a much larger fraction of energy into optical emission.

\subsection{Comparison to Galactic Magnetar SGR~1935+2154}\label{sec:magnetar}

Another potential local analog to FRB progenitors is the Galactic magnetar SGR\,1935+2154, from which FRB-like pulses have been observed \citep{Bochenek20b,CHIME20b}. Located in Galactic center, it is severely affected by dust extinction and thus despite some efforts to observe their putative optical and infrared counterparts, these have been unsuccessful \citep[e.g.,][]{De20,Zampieri22,Hiramatsu23}. However, some of these radio pulses have presented simultaneous X-rays emission \citep{Mereghetti20,Ridnaia21,Tavani21}. 

Using the simultaneous X-ray and radio detection of this source, we adopt the analysis in \citet{De20} to interpolate the expected optical-to-radio fluence ratio.  Here we assume a continuous, broadband power law between the radio and hard X-ray detections of STARE2 \citep{Bochenek20a} and its X-ray counterpart as observed by {\it Konus-Wind} \citep{Ridnaia21}.  Such a spectrum would be expected if the emission in both wavebands is dominated by a synchrotron spectrum with a peak energy at higher energies than the hard X-ray band at 18--320~keV, which is predicted by some emission models such as the synchrotron maser \citep{Metzger19,Margalit20}.  Under this assumption and the $F_{\nu} \sim \nu^{-0.46}$ spectrum predicted in \citet{De20}, we predict that the optical fluence ratio between 400~MHz and $i$-band would be $\sim3\times10^{-3}$, very close to what we predict for the Crab pulsar (Figure~\ref{fig:eta}).  Our limits can rule out such a counterpart, albeit for bursts than moderately higher radio energies (factor of $>$1000) that obtain with \frbname.

\subsection{Implications for Progenitor and Emission Models}

Finally, we compare our limits on optical counterparts to progenitor and emission models presented in \citet{Yang19}, which are shown as dashed lines in Figure~\ref{fig:eta}.  Specifically, these models correspond to emission from a pulsar magnetosphere and from a young supernova remnant (SNR), which we can rule out, as well as maser emission in an outflow from a young magnetar, which we are not able to rule out with our limits.  The first model \citep[see][]{Kumar17,Yang18} produces optical emission from energetic electrons in the magnetosphere of a pulsar, which scatter radio emission to optical wavelengths and primarily depends on the magnetic field strength and rotation rate of the young pulsar.  The former is expected to be extremely high for FRB progenitor systems \citep[e.g., SGR\,1935+2154 is $\approx$2.2$\times$10$^{14}$~G;][]{Israel16}, though the rotation period is uncertain.  For the upper range of expected fluence ratios in \citet{Yang19} (see, e.g., their Fig.~3) we can rule out such an emission mechanism.

The second emission model corresponds to inverse Compton emission from the energetic electrons in a young supernova remnant or pulsar wind nebula \citep[e.g.,][]{Piro16}.  Here the density and total energy of the electron population depends on the age, ejecta mass, and total energy of the initial explosion.  Again, we can rule out the most massive, youngest, and low energy explosions based on the range of expected values presented in \citet{Yang19}.  However, this progenitor model is complicated by the fact that extremely young SNRs would obscure the underlying FRB with free-free opacity, as well as the fact that modern time-domain surveys \citep[e.g.,][]{Bellm19,Jones21} place deep limits on the presence of a typical supernova on the timescales explored by \citet{Yang19}.

We also considered maser emission from an inverted population of electrons in an outflow or burst of ejecta around a young magnetar.  This model has been explored in detail in the literature \citep{Lyubarsky14,Beloborodov17,Waxman17,Lu18,Metzger19,Margalit20}.  It is a promising model for optical counterparts because some emission mechanisms predict a longer-lived afterglow that can in principle be detected by untargeted surveys such as the Vera C. Rubin Observatory's Legacy Survey of Space and Time \citep{Yang19} or targeted follow up of FRBs \citep{kilpatrick2021,Hiramatsu23,Trudu23}.  Given the timescale of our observations, we consider a prompt optical counterpart with a millisecond timescale, which in general will have a fluence $\approx2\times10^{-6}$ that of the radio \citep{Yang19}.  Our limits do not approach this level, leaving room for future exploration of intrinsically more energetic bursts or those much closer than \frbname. 

\subsection{Prospects for Additional High-speed Follow up of FRBs}

Given its status as one of the earliest discovered repeating FRBs, proximity at 150~Mpc, and especially its periodicity, \frbname\ has been a prime target in the search for multi-wavelength emission from FRBs \citep{Andreoni19,Pilia20,kilpatrick2021,Trudu23}.  However, the lack of detections at all frequencies a few GHz despite these concerted efforts has placed strong constraints on multi-wavelength emission counterparts and the emission mechanisms described above.  It remains open whether \frbname\ is representative of the known FRB population or if there can be multiple progenitor and emission channels with a variety of optical-to-radio fluence ratios.

\citet{Hiramatsu23} found that targeted follow up within 3~days of a new burst from a repeating FRB yielded observations coincident with a subsequent burst $\approx$40\% of the time.  As opposed to our strategy of targeting the periodic \frbname\ near the peak of its expected activity period, this strategy possible new optical burst detections across a variety of sources and deeper luminosity limits for those at closer distances \citep[e.g., the repeating FRB in a globular cluster of M81 at 3.6~Mpc, FRB\,20200120E;][]{Kirsten22}.  At the same time, untargeted follow up from optical surveys will be extremely valuable both for prompt counterpart detections \citep{Yang19} as well as pre-burst and post-burst constraints on supernova emission or more exotic optical counterparts \citep[e.g., the stellar merger counterpart in][]{Sridhar19}.  Continued optical follow up will therefore play an important role in determining the FRB mechanism and its progenitor source.

%%%%%%%%%%%%%%%%%%%%%%%%%%%%%%%%%%%%%%%%%%%%%%%%%%%%%%%%%

\section{Conclusions}\label{sec:conclusion}

We have presented high-speed ($\approx 10$\,ms) optical follow up of \frbname\ with the \alopeke\ camera at Gemini-North observatory contemporaneous with two radio bursts, \frbeventa\ and \frbeventb, detected by the CHIME array.  In summary, we find:

\begin{enumerate}
    \item There are no prompt optical counterparts in our data after correcting for the effects of dispersion, light-travel time, and the uncertainties in the internal clocks between \alopeke\ and CHIME.  Accounting for these uncertainties, we derive limits on optical fluence in each of the 10.4~ms time bins of our \alopeke\ data in $r$- and $i$-bands of $<$1.38--3.27~$\mu$Jy~s, corresponding to a total emitted optical energy of $<$8.1--32.0$\times$10$^{40}$~erg and optical-to-radio (400~MHz) fluence ratios of 2--7$\times$10$^{-3}$.
    \item Comparing to expectations for optical pulsars or the broadband optical emission from the Galactic magnetar SGR\,1935+2154, we rule out sources with the largest partition of optical-to-radio energies, which in general are around $3\times10^{-3}$.  However, there is a large range in values for these sources, such as the Vela pulsar with $2\times10^{-7}$, and limits on optical counterparts from such a source would only be possible for the closest or most energetic FRBs.  
    \item We also compare to expected models of FRBs and are able to rule out several types of inverse Compton emission presented in \citet{Yang19}, for example from a pulsar magnetosphere or supernova remnant, but not for the lowest energy inverse Compton counterparts or a synchrotron maser.
\end{enumerate}

%%%%%%%%%%%%%%%%%%%%%%%%%%%%%%%%%%%%%%%%%%%%%%%%%%%%%%%%%%%

\section*{Acknowledgements}
C.D.K. acknowledges support from a CIERA postdoctoral fellowship.
C.D.K., N.T. and J.X.P. acknowledge support
from NSF grants AST-1911140, AST-1910471, and AST-2206490
as members of the Fast and Fortunate for FRB Follow-up team.
N.T. and C.N. acknowledge support by FONDECYT grant 11191217.
\alopeke\ was funded by the NASA Exoplanet Exploration Program and built at the NASA Ames Research Center by Steve B. Howell, Nic Scott, Elliott P. Horch, and Emmett Quigley.
This work is partly based on observations obtained at the international Gemini Observatory, a program of NSF’s OIR Lab, which is managed by the Association of Universities for Research in Astronomy (AURA) under a cooperative agreement with the National Science Foundation, on behalf of the Gemini Observatory partnership: the National Science Foundation (United States), National Research Council (Canada), Agencia Nacional de Investigaci\'{o}n y Desarrollo (Chile), Ministerio de Ciencia, Tecnolog\'{i}a e Innovaci\'{o}n (Argentina), Minist\'{e}rio da Ci\^{e}ncia, Tecnologia, Inova\c{c}\~{o}es e Comunica\c{c}\~{o}es (Brazil), and Korea Astronomy and Space Science Institute (Republic of Korea). The Gemini data were obtained from programs GN-2020B-DD-103 (PI Prochaska) and GN-2022B-Q-202 (PI Prochaska).

%%%%%%%%%%%%%%%%%%%%%%%%%%%%%%%%%%%%%%%%%%%%%%%%%%%%%%%%%%%

\textit{Facilities}: Gemini (\alopeke)

\textit{Software}:
{\tt astropy} \citep{astropy},
{\tt fitburst} \citep{Fonseca23},
{\tt photutils} \citep{photutils}

\bigskip

\section*{Data and Software Availability}

All data and analysis products presented in this article are available upon request.  Analysis code and photometry used in this paper are available at \url{https://github.com/profxj/papers/tree/master/FRB/Alopeke}.  The Gemini data are publicly available on the Gemini data archive at \url{https://archive.gemini.edu/}.

 %%%%%%% ----- %%%%%%%%%%
\bibliography{frb180916_alopeke}
\bibliographystyle{aasjournal}

\end{document}